\documentclass[a4paper,11pt]{article}
\pdfoutput=1 

\usepackage{jinstpub} 

\title{\boldmath Design and Implementation of a Non-magnetic Cryogenic Multi-conductor Vacuum Feedthrough$^{\dag}$}

\author[1]{V. Cianciolo\note{Corresponding author. \newline \hspace*{4.5mm}$^{\dag}$This manuscript has been authored by UT-Battelle, LLC, under contract DE-AC05-00OR22725 with the US Department of Energy (DOE). The US government retains and the publisher, by accepting the article for publication, acknowledges that the US government retains a nonexclusive, paid-up, irrevocable, worldwide license to publish or reproduce the published form of this manuscript, or allow others to do so, for US government purposes. DOE will provide public access to these results of federally sponsored research in accordance with the DOE Public Access Plan (http://energy.gov/downloads/doe-public-access-plan).}}
\author{J.C. Ramsey}
\author{and L. Fabris}
\affiliation{Oak Ridge National Laboratory, Oak Ridge, Tennessee, USA}

\emailAdd{cianciolotv@ornl.gov}

\abstract{A common issue for experiments requiring very low magnetic fields on vacuum and cryogenic systems is how to
deal with the control of magnetic properties of electrical contacts. We describe the design and implementation of a generic sixty-four pin non-magnetic cryogenic vacuum feedthrough capable of disconnecting contacts on both sides of the vacuum boundary.}


\arxivnumber{1805.06048} 

\begin{document}

\maketitle
\flushbottom

\label{sec:intro}


The nEDM@SNS experiment~\cite{a} aims to measure the neutron electric dipole moment with unprecedented precision ($\sigma_d < 5\times10^{-28}\,{\rm e \cdot cm}$) in order to shed light on the source of charge-parity violation responsible for the generation of matter in the Universe immediately following the Big Bang~\cite{b,c}. The experiment, which is organized around the principles laid out in a seminal article by Golub and Lamoreaux~\cite{d}, is a large cryogenic apparatus with stringent non-magnetic requirements. The large number of associated sensor and control lines led to the need to develop an affordable non-magnetic, cryogenic, multi-conductor vacuum feedthrough. The design and implementation of this connector are described in this report.

\section{Design Considerations}
The nEDM@SNS experiment will have hundreds of wires (for temperature sensors, magnet coil supplies, magnetic field sensors, heaters, etc.) that need to penetrate the outer room-temperature vacuum boundary. Many of these wires will also need to penetrate an inner cryogenic (4K) vacuum boundary. The experiment's non-magnetic requirements (<100\,nT on contact\footnote{Components are measured with a fluxgate magnetometer before and after exposure to a strong horseshoe magnet.} for components mounted on the cryostat) make most commercially available connectors unacceptable due to the nearly ubiquitous presence of a nickel flash used to prevent diffusion of the base conductor material through the typical gold or tin anti-corrosion coating. 

Having concluded a custom connector was required, our goal was to design a generic connector useful for most, if not all, the experiment's electrical vacuum penetration needs. Another goal was to retain the ability to disconnect signals from either side of the vacuum boundary without disturbing the vacuum boundary. This simplifies assembly, adds flexibility and reduces the likelihood of introducing a new leak in a previously leak-tight system (an essential feature given the size of the system and the associated long turnaround times). This is not the first published cryogenic non-magnetic multi-conductor connector (see~\cite{de0}). The connector described in this article required more engineering and is probably more expensive per contact, but does feature a glue-free vacuum seal, requires less custom assembly and allows simultaneous connection or disconnection of all contacts - all helpful features given the number of contacts in the nEDM@SNS experiment.

\section{Design and Implementation Details}
\begin{figure}[htbp]
\centering 
\includegraphics[width=6in]{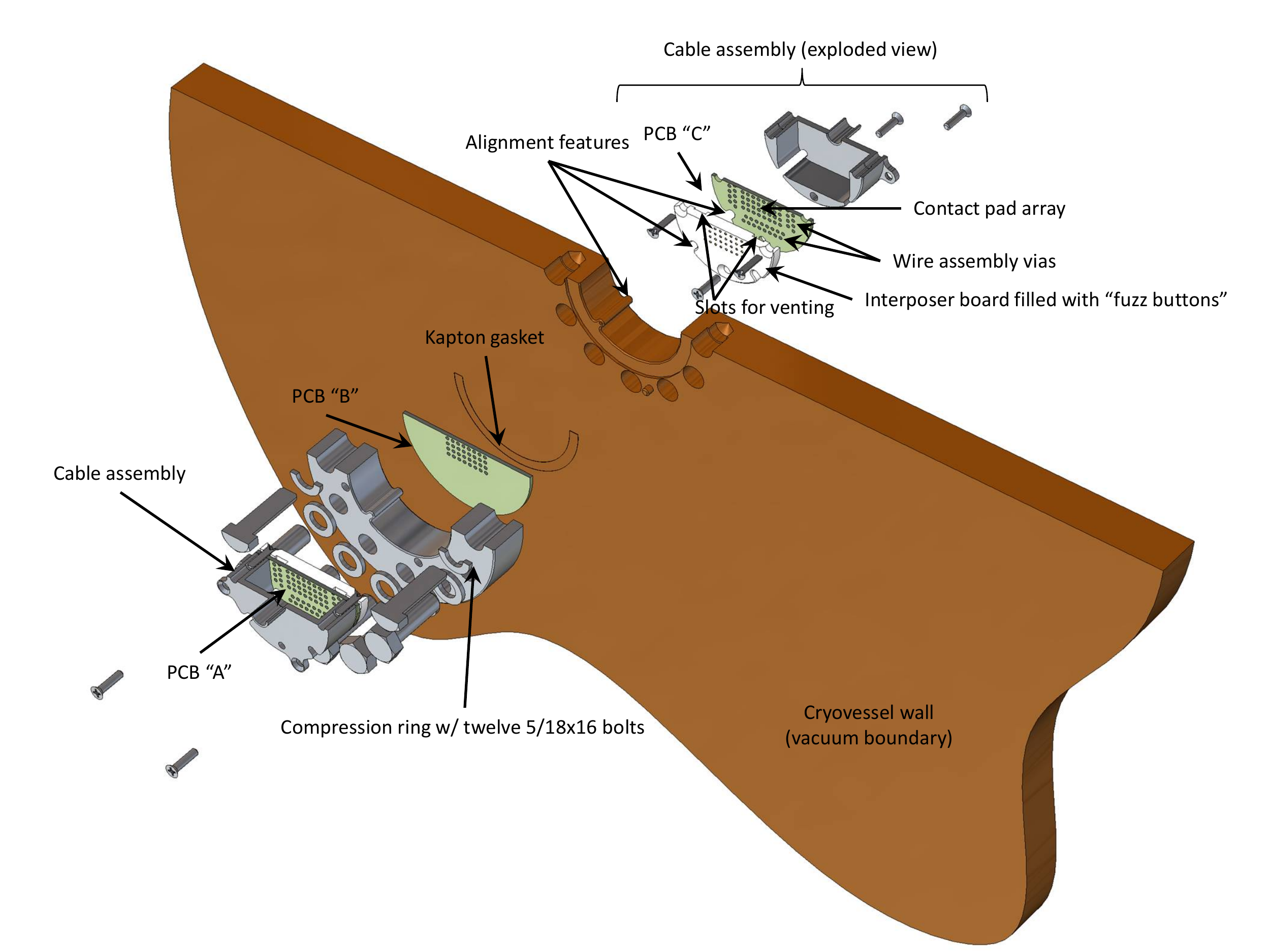}
\caption{\label{fig:1} Cut-away view of the connector; details in the text.}
\end{figure}

\begin{figure}[htbp]
\centering 
\includegraphics[width=6in]{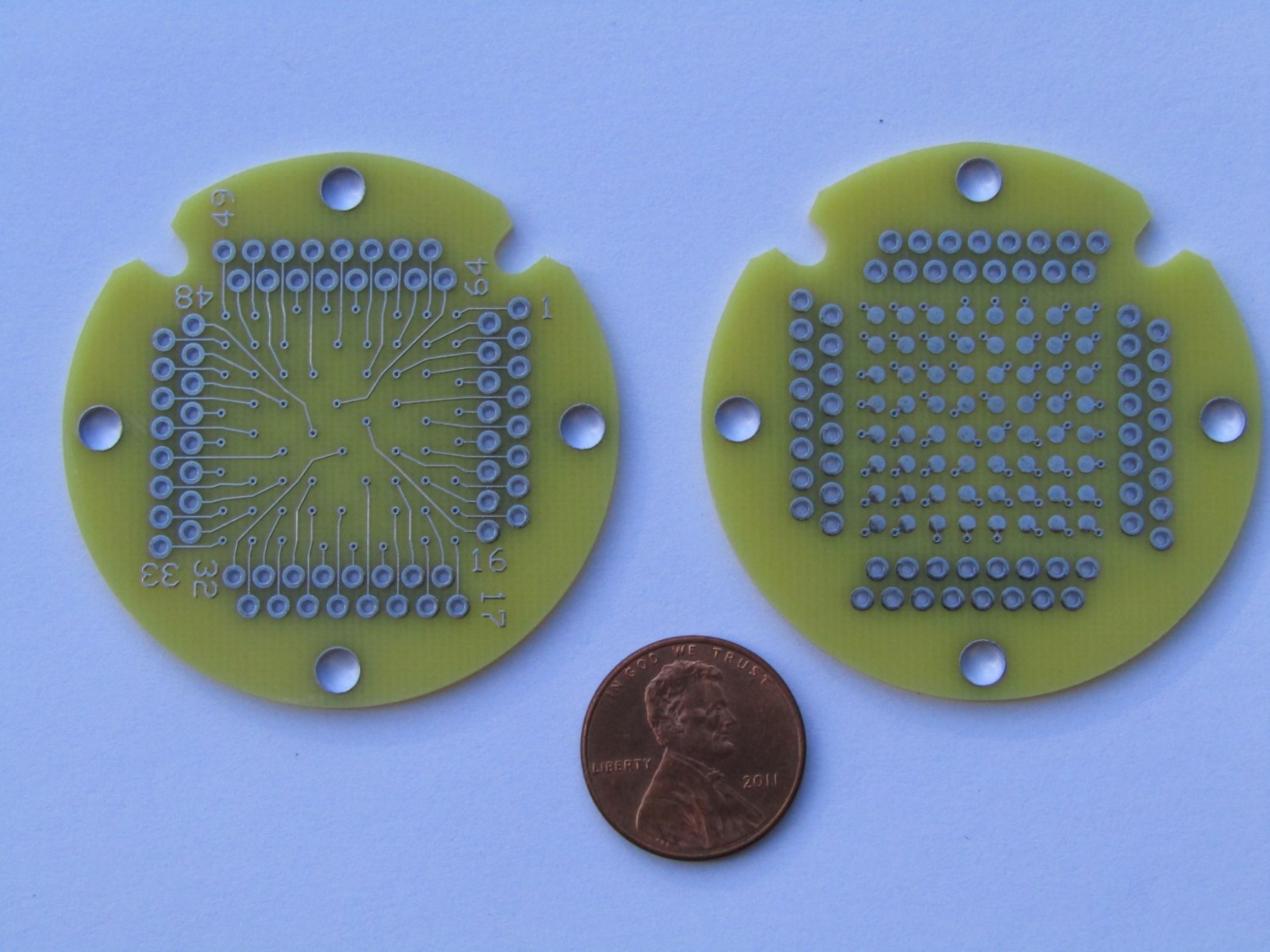}
\caption{\label{fig:pcba} The two sides of PCB "A" - the input connector. Signal wires are soldered into the receptacles on the outer edge. Traces on the top layer (left-hand image) - carry the signals to vias which bring the signals to the opposite side of the board (right-hand image) which has a pad array and no traces (which can short-circuit to neighboring fuzz buttons). The pad array is compressed onto an interposer board, shown in Figure~\ref{fig:interposer}. PCB "C" - the output connector - is identical except for the alignment features (the notches at the outer diameter).}
\end{figure}

\begin{figure}[htbp]
\centering 
\includegraphics[width=6in]{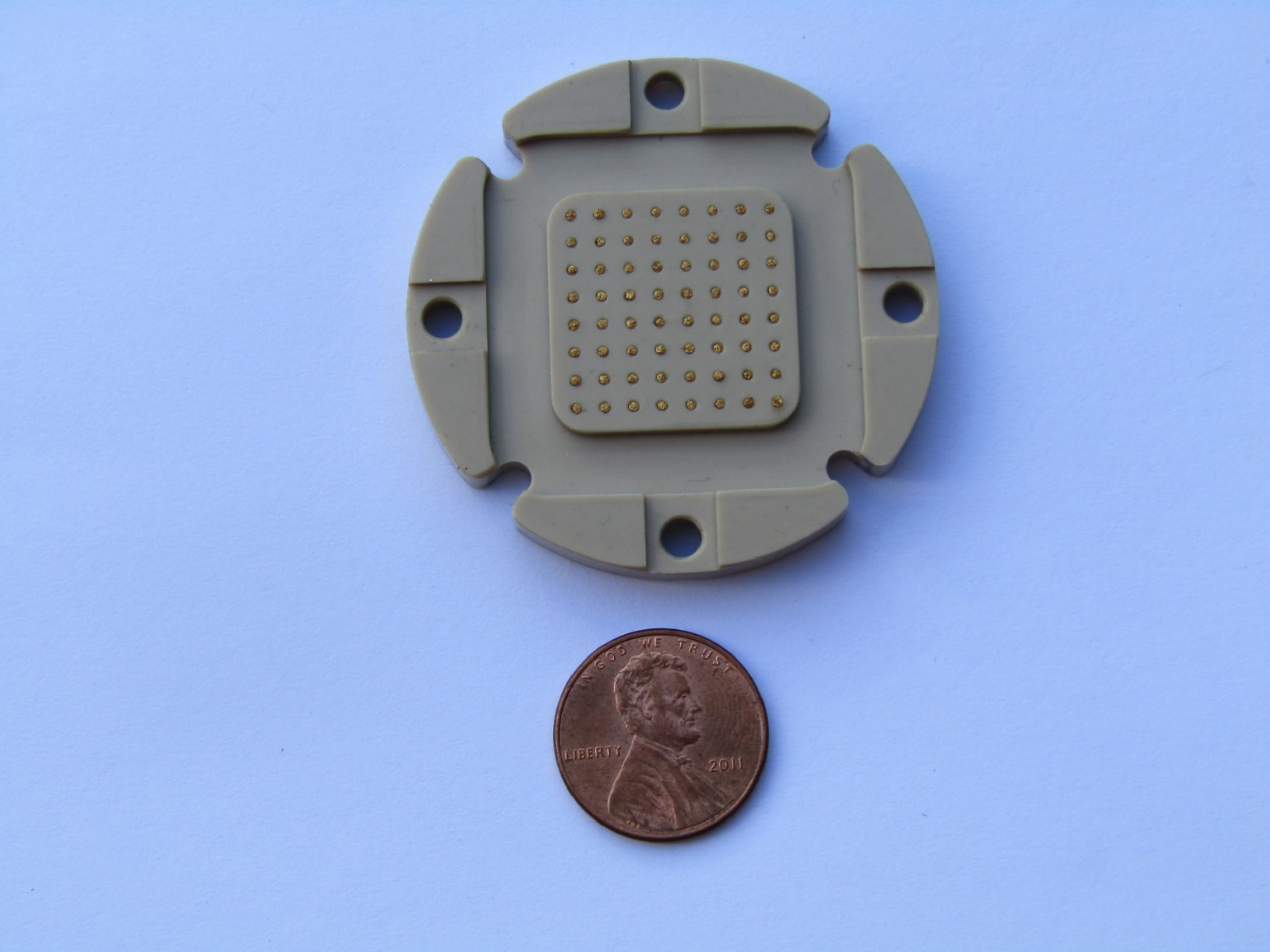}
\caption{\label{fig:interposer} View of the PEEK interposer board with its array of fuzz buttons. Slots on the side prevent virtual leaks and provide clearance for solder connections of the signal wires. Interposer boards for the input and output connector are identical.}
\end{figure}

Figure~\ref{fig:1} shows the basic concept of the connector described in this report. Wires on both sides of the vacuum boundary are soldered into receptacles on the outer edge of a printed circuit board ("A" or "C"). Traces bring the signals on each printed circuit board (PCB) from the wire receptacles to a set of vias that bring the signals to a pad array on the other side of the board. The two sides of PCB "A" are shown in Figure~\ref{fig:pcba}. Interposer boards (shown in Figure~\ref{fig:interposer}) mounted on PCB "A" and PCB "C" host a series of pressure-enabled contacts ("fuzz buttons", described below) that are aligned with the corresponding pad arrays. Once assembled, the wires, PCBs and interposer boards form two cable assemblies that are plugged into PCB "B" to make the necessary electrical connections across the vacuum boundary. This achieves the goal of being able to disconnect signals from either side of the vacuum boundary without disturbing the vacuum boundary.

\begin{figure}[htbp]
\centering 
\includegraphics[width=6in]{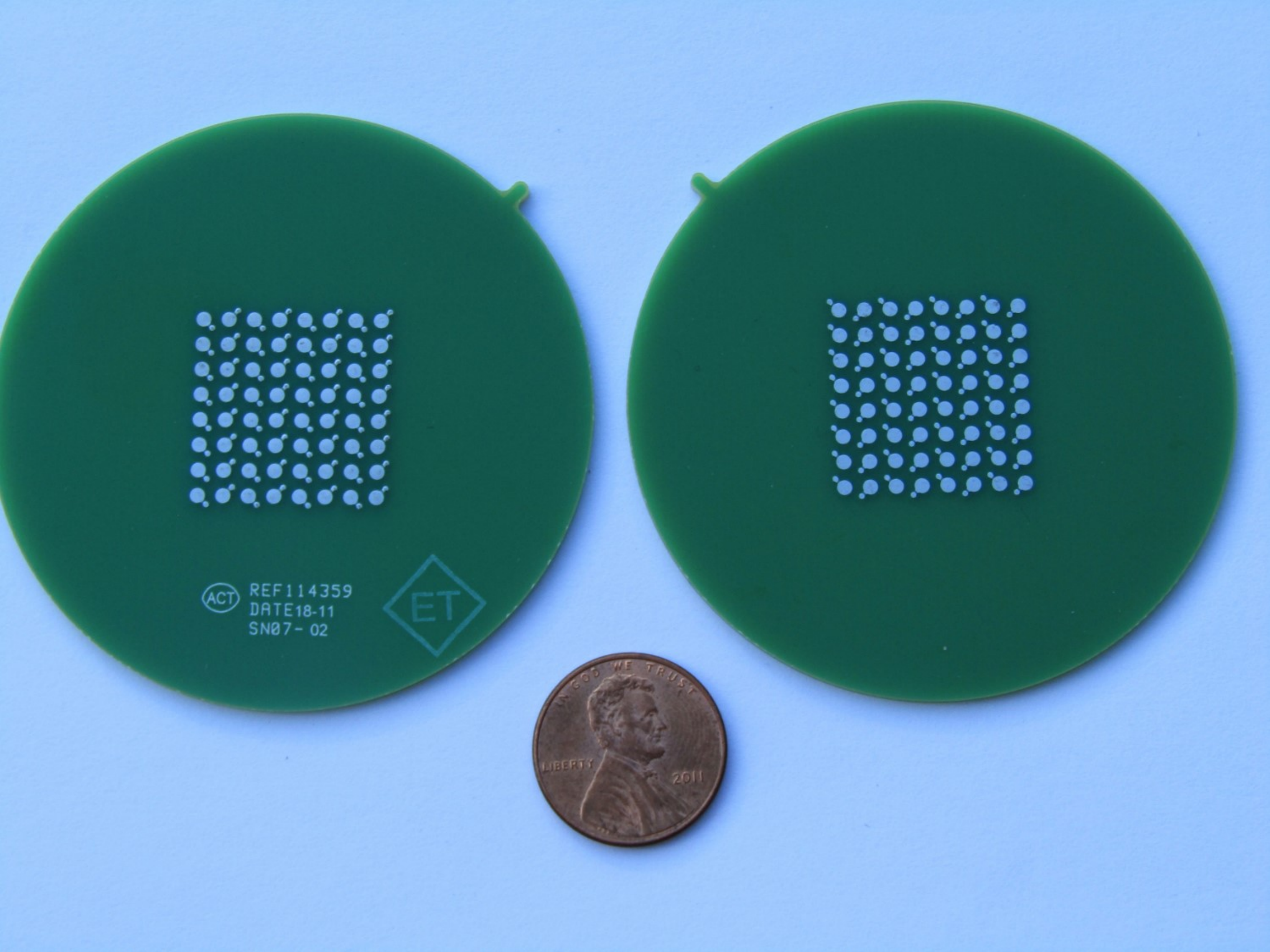}
\caption{\label{fig:pcbb} The two sides of PCB "B", which forms part of the vacuum boundary. Identical pad arrays on the two sides of the board are electrically connected with vias - specifications listed in the text. The interposer boards from the input and output connectors are brought into contact with pads on the opposite sides of PCB "B" to establish the electrical connection across the vacuum boundary.}
\end{figure}

PCB "B" (Figure~\ref{fig:pcbb}) forms part of the vacuum boundary. PCB "B" is a two-layer 0.062"-thick 370HR board, fabricated to IPC 6012 Class DS~\cite{de} specifications by Advanced Circuits. 370HR is an FR-4-based material and FR-4 materials generally have good outgassing properties~\cite{de2}. The two signal layers are 1\,oz copper, tinned with SN63 alloy.  Sixty-four capped and filled vias (24 mil O.D., 8 mil O.D. drill size) bring the signals from the top signal layer to the bottom, thus crossing the vacuum boundary. The boards were found to have a small ($\leq 20$\,nT) magnetic signal on contact (below our upper limit). To complete the vacuum boundary an annular kapton gasket (2.3" O.D., 2.0" I.D., 5 mil thick) is compressed between PCB "B" and the cryostat body using an aluminum compression ring. The sealing surface is finished to 16\,$\mu$-inch. The compression force ($\approx$ 16,300\,psi)  applied by twelve $5/16-18$ aluminum bolts torqued to 90\,inch-lbs exceeds the suggested minimum compression force~\cite{e} by more than a factor of two. Titanium helicoils are used to avoid stripping threads in the aluminum cryostat body while satisfying non-magnetic requirements. The compression ring thickness (0.66") was chosen based on experience with indium-sealed and conflat flanges of similar size, but oversized to enable use of easily available aluminum bolts and for compatibility with the cable end component stack.

"Fuzz buttons"~\cite{ef} from Custom Interconnects provide electrical contact between mating pad arrays on two vertically stacked PCBs. Fuzz buttons are tiny cylindrical bundles of wire. To make electrical contact, pads on the PCBs that are to be mated are mutually aligned with a fuzz button secured in an interposer board (Figure~\ref{fig:interposer}), and the PCB-interposer-PCB sandwich is squeezed together (features on all three boards and on the connector housing ensure proper alignment). As a result there is no requirement for pins to penetrate the vacuum boundary. Fuzz buttons are available in gold-plated beryllium-copper wire (Au/BeCu), with copper instead of nickel flash under the gold plating; a formulation found to have no detectable magnetic signal (again using a fluxgate magnetometer following exposure to a strong magnetic field). Contact resistance varies with the relative compression of the fuzz button; for 30\,mil Au/BeCu fuzz buttons 40-50\,m$\Omega$ contact resistance is obtained with 15-20\% compression, corresponding to $\approx$0.1\,lbs of applied force~\cite{ef1}. Fuzz buttons come in a variety of lengths and widths and can be arranged in fully-customizable two-dimensional patterns. 

Custom Interconnects has published design rules for both interposer boards that hold the fuzz buttons and the mating PCBs. Nothing beyond those design rules was required for successful cryogenic operation, although we did choose PEEK for the interposer (from among the various materials suggested) due to positive experience with this material in other cryogenic applications. Appropriate slots were added to the interposer to ensure no virtual leaks during vacuum operation.

We chose sixty-four contacts per connector as a balance between minimizing the number of total connectors and minimizing the diameter of the kapton seal needed for PCB "B". We chose 30\,mil diameter fuzz buttons so that each contact was capable of carrying 6\,A of current (any circuit requiring more current can be split onto multiple parallel contacts).  

To make good electrical contact for sixty-four 30\,mil fuzz buttons 7.7 lbs of compressive force is required (provided by four 4-40 bolts). We chose a 100\,mil pad pitch and 60\,mil pad diameter to make inter-board alignment straightforward, albeit with careful machining and precision alignment pins. The resulting 40\, mil pad separation allows 300\,V differential on adjacent pads (Milspec 275D specifies required PCB conductor spacing of 0.12\,mil/V). 

\section{Performance}
Using an environmental chamber a sample connector was subjected to seven thermal cycles (from room temperature down to 123\,K, limited by the test chamber minimum temperature) with $dT/dt \approx 15$\,K/minute). At the lowest temperature the thermal contraction of all components is roughly 80\% of the asymptotic ($T=0$) value~\cite{f}. One side of the contact was exposed to vacuum (sampled by a Helium leak tester) for the duration of the test, the other side was flooded with helium gas. No leak was observed (backgrounds < 10$^{-10}$\,mbar$\cdot$l/s). We did not perform a superfluid leak test, but our anticipated applications for this connector are all above 4\,K. Electrical connection was monitored during the entire temperature cycle. Contact resistivity was measured with a four-point sensor; minimal increase in contact resistivity ($<3\%$) was observed. High-frequency response was tested by injecting TTL pulses; no degradation of the signal shape was observed.

The materials cost of the connector is ~\$10/contact, dominated by the two fuzz buttons per contact.

\section{Conclusions}
We have developed an affordable non-magnetic, cryogenic, multi-conductor vacuum feedthrough. This connector was developed explicitly for the nEDM@SNS experiment, but the design is generic enough that it may find use in other similar applications.

\acknowledgments

We gratefully acknowledge the support of the U.S. Department of Energy Office of Nuclear Physics through
contract DE-AC05-00OR22725. We would also like to thank John Whiteley and Carrie Wilson at Advanced Circuits for expert assistance with the via design for the vacuum boundary PCB and Dr. Simon Slutsky at Caltech for magnetic measurements of the components.



\end{document}